\pdfoutput=1
\documentclass{article}

\usepackage{arxiv}
\makeatletter
\renewenvironment{table}
  {\setlength{\abovecaptionskip}{\@abovecaptionskip}%
   \setlength{\belowcaptionskip}{\@belowcaptionskip}%
   \@float{table}}
  {\end@float}
\makeatother
\usepackage[T1]{fontenc}
\usepackage{amsmath}
\usepackage{amssymb}
\usepackage{amsfonts}
\usepackage{graphicx}
\usepackage{booktabs}
\usepackage{float}
\usepackage{nicefrac}
\usepackage{microtype}
\usepackage{natbib}
\usepackage{xspace}
\usepackage{fontawesome5}
\usepackage{tabularx}
\usepackage{longtable}
\usepackage{enumitem}
\setlist[itemize]{itemsep=2pt, topsep=4pt, parsep=0pt}
\usepackage{tikz}
\usetikzlibrary{arrows.meta,positioning,calc}
\usepackage{pgfplots}
\pgfplotsset{compat=1.17}
\definecolor{seriesA}{HTML}{2A78D6}
\definecolor{seriesB}{HTML}{EB6834}
\definecolor{seriesC}{HTML}{1BAF7A}
\definecolor{seriesD}{HTML}{EDA100}
\definecolor{linkblue}{HTML}{0969DA}
\definecolor{seriesE}{HTML}{E87BA4}
\definecolor{seriesF}{HTML}{008300}
\definecolor{seriesG}{HTML}{6250D6}
\colorlet{solvedbg}{black!6}
\colorlet{stepsfill}{seriesA!32}
\usepackage{colortbl}
\newcommand{\ci}[2]{{\scriptsize\textcolor{black!55}{[#1, #2]}}}
\usepackage{hyperref}
\usepackage{url}
\usepackage{doi}
\usepackage{cleveref}

\graphicspath{{figures/}}

\newcommand{\bench}{Multi-step SWE\xspace}

\title{Multi-step SWE: Can Coding Agents Keep Building on Their Own Code Over Long Horizons?}

\author{%
  Xiangkai Zeng\thanks{Equal contribution; authors are listed in alphabetical
  order by first name.} \qquad
  Zheng Liu$^{\ast}$ \qquad
  Zihao Lin$^{\ast}$ \\
  Klavis AI \\
  \texttt{\{xiangkaiz,\,zhengliu,\,zihaolin\}@klavis.ai} \\
}

\date{}

\renewcommand{\shorttitle}{}

\renewcommand{\headeright}{}
\renewcommand{\undertitle}{}

\hypersetup{
  hidelinks,
  pdfpagelayout=OneColumn,
  pdftitle={Multi-step SWE: Can Coding Agents Keep Building on Their Own Code Over Long Horizons?},
  pdfsubject={cs.SE},
  pdfauthor={Xiangkai Zeng, Zheng Liu, Zihao Lin},
  pdfkeywords={software engineering, coding agents, benchmark, long-horizon evaluation, multi-step tasks},
}

\begin{document}
\maketitle

\begin{abstract}
  Developers rarely give a coding agent a single request: they ask it to build
  a component and then to extend, upgrade, and optimize it, each time on code
  the agent itself wrote. Most benchmarks, however, score one task at a time.
  We introduce \bench, a long-horizon benchmark of 69 software engineering
  tasks, each a chain of 5--15 dependent steps on a single codebase. Every step
  is a complete, multi-hour engineering request that starts from the code the
  agent produced in the preceding steps, so early design decisions and errors
  carry forward. The tasks span four families of engineering work and nine
  languages, include both from-scratch builds and existing open-source
  codebases, and ground every step in an existing source of truth that defines
  both the instruction and its verifier. Across nine frontier models, the
  strongest model reaches a solve rate of only 18.5\%, and 38 of the 69 tasks
  are not solved by any model. We release \bench publicly.
\end{abstract}

% Project links: a trophy (Noto Emoji, Apache 2.0) and the GitHub mark, with
% blue link text.
\begin{center}
  \href{https://multistepswe.com/}{\raisebox{-0.2\height}{\includegraphics[height=1.15em]{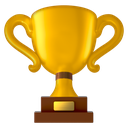}}~\textcolor{linkblue}{Leaderboard}}%
  \hspace{2.5em}%
  \href{https://github.com/Klavis-AI/Multi-step-SWE}{\faGithub~\textcolor{linkblue}{GitHub}}
\end{center}

% ===========================================================================
\section{Introduction}
\label{sec:intro}

Coding agents have become part of everyday software development, and
developers increasingly rely on them for long-horizon work that unfolds over a
series of requests. A developer may ask an agent to build a component, then to
extend it, to port it to a new framework release, and finally to accelerate a
performance-critical path. Each request operates on code shaped by the earlier
requests, frequently code that the agent itself wrote, and many requests
require hours of agent work. The success of the final request therefore
depends both on the quality of the first three and on the agent's ability to
understand, extend, and preserve the existing code.

\begin{figure}[t]
  \centering
  \includegraphics[width=\linewidth]{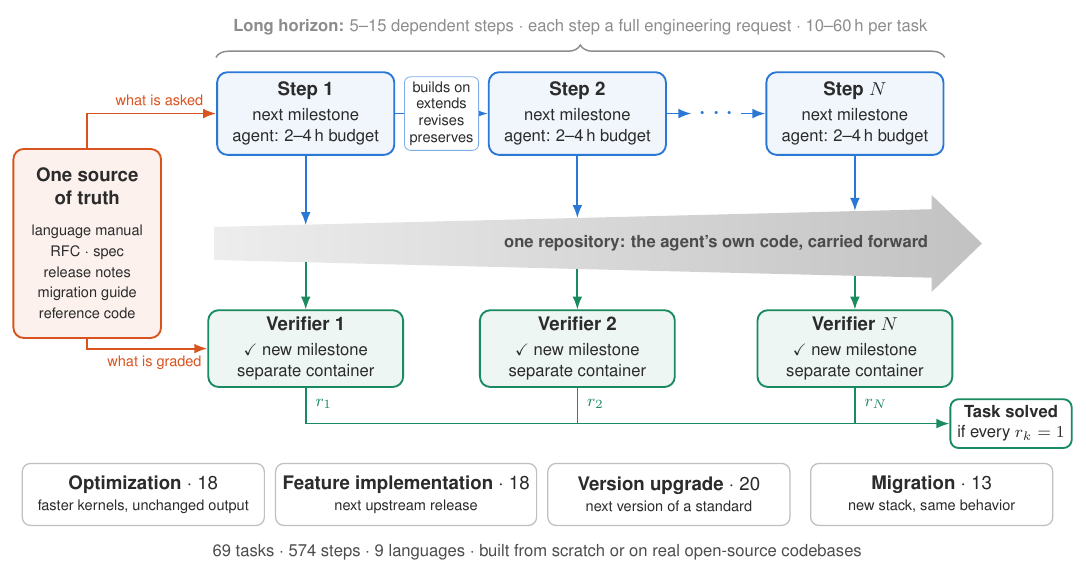}
  \caption{Overview of \bench. A task chains dependent steps on one
    repository. Each step is a complete engineering request that starts from the code the
    agent produced in earlier steps and is graded by a dedicated verifier. A
    task is solved only if every step receives full credit. A single source of
    truth defines both the requirements of each step and the way the step is
    graded.}
  \label{fig:overview}
\end{figure}

Most benchmarks still evaluate one task at a time. SWE-bench asks a model to
resolve a single GitHub issue \citep{jimenez2023swebench}; Terminal-Bench,
DeepSWE, and FrontierSWE pose harder and longer tasks of up to 20 hours each
\citep{merrill2026terminalbench,huang2026deepswe,chu2026frontierswe}; and
Commit0 and ProgramBench ask agents to build an entire library or program from
scratch \citep{zhao2024commit0,yang2026programbench}. Regardless of length,
each of these tasks is a single episode that is graded once at the end, so none
of them requires an agent to keep building on its own earlier work as the
requirements change. Recent sequential benchmarks chain dependent changes,
such as package upgrades, feature checkpoints, and milestones mined from
repository history
\citep{orlanski2026slopcodebench,lam2026swechain,shen2026evocodebench,deng2026swemilestone,shastry2026swesteps},
and they report that the performance of agents degrades as work accumulates.

We introduce \bench, a multi-step, long-horizon benchmark of 69 software
engineering tasks in which a coding agent carries one codebase through 5--15
dependent steps, each built on the agent's own earlier code
(\Cref{fig:overview}). It extends the sequential setting in three directions:
in the scale of each step, which is a complete engineering request with a
budget of several hours; in grounding, since every step is tied to an existing
source of truth and graded by the field's own tools; and in breadth, with nine
languages and four families of engineering work, including performance
optimization and stack migration.

\textbf{A step is a complete engineering request, not an agent turn or a tool
call.} Each step corresponds to a request that a developer might delegate to an
agent, such as ``move the interpreter to Lua~5.4'' or ``make the AES kernels
faster,'' and has its own instruction, verifier, and time budget of 2--4 hours.

\textbf{Each step builds on the agent's own code.} Step $k$ starts from the
workspace that the agent left after step $k-1$ rather than from a reference
solution, so the agent's design decisions and errors carry forward. This
property distinguishes \bench from a collection of independent tasks: every
step after the first is attempted on code that the agent itself has written or
modified.

Coding benchmarks are susceptible to instruction--verifier mismatch, in which a
verifier checks details that the instruction does not state or rejects valid
solutions \citep{openai2024swebenchverified}. \bench mitigates this problem by
grounding every step in an existing source of truth that already defines the
expected behavior, such as specifications, upstream release histories, the
behavior of an original codebase, and reference implementations. Each step
provides or cites this material, and the verifier grades against it with the
established tools of the field, so that each instruction and its verifier are derived from the same
material (\Cref{sec:alignment}).

We evaluate nine frontier models with four rollouts per task and consider a
rollout solved only if every step receives full credit. Whole tasks are seldom
solved. The strongest model, Claude Opus 5.5, reaches a solve rate of 18.5\%
and solves 31.9\% of the tasks within four attempts; every other model has a
solve rate of at most 17.8\%, and more than half of the tasks are not solved by
any model.

Our contributions are as follows:
\begin{itemize}
  \item \textbf{A broad multi-step, long-horizon benchmark.} \bench spans four
    families of engineering work (performance optimization, feature implementation, version
    upgrade, and migration), nine languages, and both from-scratch builds and
    existing open-source codebases, ranging from interpreters and databases to
    network protocols and cryptography. All tasks are packaged in the open
    Harbor multi-step task format.
  \item \textbf{Well-defined steps with aligned instructions and verifiers.}
    Every step is grounded in an existing source of truth and graded at the
    artifact's external interface by the field's own conformance suites,
    reference tools, recorded baseline behavior, or speedup measurements
    gated on unchanged output, so that what is requested and what is graded are designed to
    agree. Every task passed two rounds of review by expert software
    engineers with at least five years of experience, and every step has a
    reference solution.
  \item \textbf{An evaluation of nine frontier models.} The evaluation shows
    that models often pass individual steps but rarely complete whole tasks,
    that failing agents often stop
    early and report success, having checked their work only against their own
    expectations.
\end{itemize}

% ===========================================================================
\section{The \bench Benchmark}
\label{sec:benchmark}

\subsection{Overview}
\label{sec:principles}

\bench is designed to measure the capability of coding agents in long-horizon
development, where a sequence of substantial requests is carried out on code
that the agent itself wrote. To this end, it follows four design principles
(\Cref{tab:principles}). Tasks are \emph{multi-step}, so that the evaluation
reflects how developers actually use coding agents. Each step is a
\emph{long-horizon} engineering request, so that the benchmark is challenging
for frontier models. Each step is \emph{well defined}, so that the tasks are of
high quality and are graded accurately. Finally, the tasks are \emph{diverse},
so that they test agents along different dimensions of software engineering.

\begin{table}[h!]
  \centering
  \small
  \begin{tabularx}{\linewidth}{@{}l X@{}}
    \toprule
    Principle & Realization in \bench \\
    \midrule
    Multi-step & Each task chains dependent steps on one codebase, and each
      step starts from the code the agent produced in the preceding
      steps. \\
    Long horizon & Each step is a complete engineering request with a budget of
      several hours; reference solutions reach tens of thousands of lines of
      code. \\
    Well defined & Each step is grounded in an existing source of truth that is
      provided in the workspace or cited in the instruction, and its verifier
      uses the field's own test suites and reference tools; every task passed two rounds of expert
      review. \\
    Diverse & Four task families, nine languages, 34 tasks built from scratch,
      and 35 tasks on existing open-source codebases. \\
    \bottomrule
  \end{tabularx}
  \caption{Design principles of \bench and how each principle is realized.}
  \label{tab:principles}
\end{table}

Every task is packaged in Harbor's multi-step task format \citep{harbor2026},
in which each step has its own instruction, verifier, and reference solution,
and can add files to the workspace (\Cref{sec:protocol} describes the format and the execution
protocol in detail). The steps run in order on a single cumulative workspace, each within a fixed wall-clock budget. After each step, the workspace is graded in a separate verifier
container, and every step is scored even if an earlier step received a reward
of zero.

The remainder of this section describes the four task families, the relations
between consecutive steps, the measures that keep instructions and verifiers
aligned, the verification and reward scheme, the defenses against reward
hacking, and the composition of the benchmark.

\subsection{Task Families}
\label{sec:families}

The tasks fall into four families (\Cref{tab:families}), defined by what each
step asks of the agent and by how previously correct behavior is treated. In
every family, a step gives the agent the workspace left by the previous step,
an instruction, and any new material copied into the workspace, while the
verifier and its tests remain hidden. Every step must also keep earlier
behavior working.

\begin{table}[h!]
  \centering
  \footnotesize
  \begin{tabularx}{\linewidth}{@{}>{\raggedright\arraybackslash}p{2.0cm} *{4}{>{\raggedright\arraybackslash}X}@{}}
    \toprule
    Family & Starting workspace & Each step provides & Each step must & Example task \\
    \midrule
    Optimization (18)
      & a real library with the optimized code of the scored kernels removed
      & the scored functions and a guide to build, check, and time them
      & make the next kernel family faster, with unchanged output
      & BoringSSL, one cryptographic kernel family per step \\
    \addlinespace
    Feature implementation (18)
      & empty, a skeleton, or an earlier release of the project
      & an interface contract and, usually, the release's specification
      & implement the capabilities of a later upstream release
      & a Tor client in Rust across 12 releases \\
    \addlinespace
    Version upgrade (20)
      & the first version's specification, with no implementation
      & the next version's specification and, usually, sample tests
      & implement the next version, changing behavior where it changes
      & a Lua interpreter in Java, from Lua 5.1 to 5.5 \\
    \addlinespace
    Migration (13)
      & a legacy codebase at a pinned revision
      & an instruction naming the target stack and the required changes
      & move to the next stack release, keeping behavior unchanged
      & CityVoice, from Rails 4.0 to 8.1 \\
    \bottomrule
  \end{tabularx}
  \caption{The four task families: where a task starts, what each step provides
    to the agent, what each step must achieve, and one example task.}
  \label{tab:families}
\end{table}

\paragraph{Optimization (18 tasks).}
The libraries include BoringSSL, simdutf, and FFmpeg; with their own optimized
kernels removed, they are correct but slow. Outputs must stay unchanged:
bit-exactly in 16 tasks, and within a tolerance stated in the task for the other two (FFmpeg and LAMMPS). The verifier checks the outputs and
measures the speedup on its own inputs.

\paragraph{Feature implementation (18 tasks).}
Each task follows the release history of a real project, such as a Tor client
or the turso database engine. The interface contract names the graded entry
points and their expected behavior, and the verifier runs hidden tests through
it, often comparing outputs with a reference implementation such as stock
\texttt{sqlite3}.

\paragraph{Version upgrade (20 tasks).}
Each task carries a standard or tool through successive versions, such as
Lua~5.1 to 5.5 or the Ethereum hard forks from Frontier to Osaka. The verifier
runs the version's official test suite, such as PHP's \texttt{.phpt} tests, or
compares outputs with those recorded from a reference implementation.

\paragraph{Migration (13 tasks).}
Each task moves a legacy open-source codebase, such as a Rails application, one
stage forward per step. The instruction also names the legacy constructs to
remove, and the code must genuinely use the new stack, with deprecated APIs
removed rather than silenced. The verifier replays hidden recordings of the
original codebase's behavior, adjusted for documented changes, and applies
static checks with standard tools such as linters.

\subsection{Relations Between Consecutive Steps}
\label{sec:relations}

\begin{figure}[h!]
  \centering
  \includegraphics[width=\linewidth]{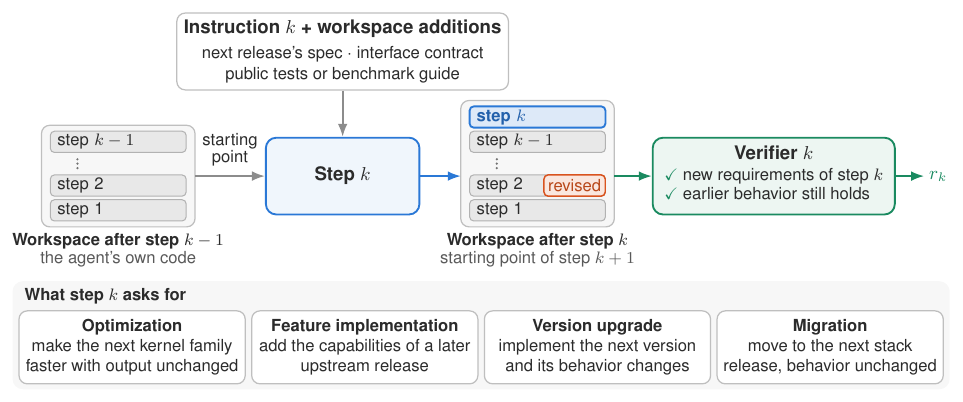}
  \caption{Relation between consecutive steps. Step $k$ starts from the
    workspace that step $k-1$ left and receives its own instruction and
    workspace additions. It extends or revises the code as its task family
    requires, and its verifier checks the new requirements of step $k$ and,
    in most tasks, re-tests behavior inherited from earlier steps.}
  \label{fig:relations}
\end{figure}

The steps of a task are not arbitrary partitions of one large task. Each step
corresponds to a real unit of progress in the underlying project: the next
version of a standard, the next upstream release of a project, the next stage
of a migration, or the next kernel family to optimize. Step $k$ is tied to the
preceding steps in two ways (\Cref{fig:relations}). It starts from the code
that steps $1,\dots,k-1$ produced, and its requirements are stated relative to
that code, as a change to make on top of behavior that must be kept.

\paragraph{Shared workspace.}
Step $k$ begins from exactly the workspace in which step $k-1$ ended, and its
instruction describes the target relative to that workspace rather than from
scratch. For example, the instruction for the fourth step of the Lua task
begins: ``This workspace holds a Lua implementation written in Java. Make it
implement Lua~5.4 instead of the release it implements now.'' Likewise, each
migration step operates on the source tree produced by the previous migration
step.

\paragraph{What step $k$ changes.}
The new work in step $k$ differs by family. In feature implementation and
version upgrade tasks, step $k$ adds capabilities that rely on components built
earlier: in the turso task, index creation and maintenance are introduced only
after the pager, B-tree, and write-ahead log of the first step are in place. In
optimization tasks, step $k$ speeds up the next kernel family of the same
library, often following the library's dataflow (FFmpeg's scaling pipeline) or
its layering, so that later kernels call earlier ones (BoringSSL's AES-GCM
runs its counter mode on the AES block cipher of the first step, and
AES-GCM-SIV reuses the GHASH code of the second).

Some steps also revise behavior that an earlier step required. This happens
only where the step's source of truth changes that
behavior, and the verifier then expects the new result. For example, Lua~5.3
introduces an integer subtype that changes numeric results, a tree-sitter
release reverses a capture-precedence rule, a turso release fixes a defect
present in earlier releases, and a move to a newer Rails release changes the
responses to certain requests. Optimization steps never revise behavior, since
their outputs must stay unchanged.

\paragraph{What step $k$ keeps.}
Everything else that earlier steps established must keep working after step
$k$: whatever the new release or version leaves unchanged in feature
implementation and version upgrade tasks, the original codebase's behavior in
migration tasks, and the unoptimized library's output in optimization tasks.
Most verifiers re-test this behavior alongside the
new requirements, so breaking earlier work generally
costs the step its reward.

\subsection{Aligning Instructions and Verifiers}
\label{sec:alignment}

An instruction--verifier mismatch occurs when a verifier checks behavior that
the instruction does not specify, or when the instruction permits solutions
that the verifier rejects. Such mismatches are a known source of error in
coding benchmarks: in the annotation effort that produced SWE-bench Verified,
38.3\% of the sampled SWE-bench tasks were flagged for underspecified problem
statements and 61.1\% for tests that may reject valid solutions
\citep{openai2024swebenchverified}. Their effect
compounds over a long task because every step adds requirements. \bench guards
against mismatches in four ways.

\paragraph{A single source of truth for instruction and verifier.}
The target behavior of each step is defined by material that exists
independently of the benchmark, which the step provides in the workspace or
cites in its instruction. Wherever
possible, the verifier is built from the same material, so that it tests what
the source of truth specifies rather than the behavior of one particular
solution. The sources are of three kinds:
\begin{itemize}
  \item \textbf{Specifications} of languages, protocols, and file formats,
    including RFCs and PEPs (Lua~5.1 to 5.5, Kubernetes~1.25 to 1.33, QUIC from
    an IETF draft to RFC~9000). The verifier runs the accompanying conformance
    material, such as the official Lua test suites, php-src's \texttt{.phpt}
    tests, Samba's \texttt{smbtorture}, or BoringSSL's BoGo suite.
  \item \textbf{Upstream documentation, changelogs, and release notes} (15
    releases of the turso database engine). The verifier tests the documented
    behavior, for example by comparing outputs with a reference implementation
    such as stock \texttt{sqlite3}.
  \item \textbf{The original codebase}, together with the target stack's
    upgrade guides and release notes (Rails~4.0 to 8.1, Spark~2.4 to 3.4). The
    verifier replays behavior recorded from the original codebase, adjusted
    only where the instruction documents a change.
\end{itemize}

\paragraph{Explicit contracts at the external interface.}
Each instruction specifies the interface that the verifier uses and the outputs
that must match the reference. Verifiers check behavior only through that
interface (the artifact's command line, API, wire protocol, or file format), so
correct solutions are not penalized for their internal structure. The few
checks that inspect source code enforce rules that the instruction states
explicitly, such as the implementation language or the removal of a legacy
API.

\paragraph{Public checks, hidden grading.}
Many steps also give the agent public checks that it can run while it works,
such as sample conformance programs with a test runner, or commands that
compare outputs with the reference implementation and time the scored
functions on practice inputs. The verifier grades the step on its own test
cases or inputs, which in most tasks are hidden from the agent and differ from
the public checks; some verifiers also rerun public tests, such as an official
conformance suite. The public checks let the agent confirm its reading of the
instruction, while the hidden cases prevent a solution from being tuned to the
graded workload.

\paragraph{Expert review.}
Every task was reviewed in two rounds by expert software engineers, each with
at least five years of professional experience. The initial
review examines the design: whether each step is a real unit of progress,
whether the instruction and specification state everything that the step
grades, and whether the step fits its time budget. The final review compares
each instruction with its verifier and confirms that every graded behavior is
either stated explicitly or follows from the source of truth, and that the
verifier cannot be passed without completing the work. Identified problems are
corrected and re-reviewed, and a task is admitted to the benchmark only after
it passes the final round.

\subsection{Verification and Reward}
\label{sec:verification}

Each step has its own verifier, which is built from that step's test directory
and runs in a separate container on a copy of the workspace produced by the
agent.

\paragraph{Step reward.}
For feature implementation, version upgrade, and migration steps, each verifier
checks what step $k$ newly requires, such as the capabilities of the new
release, the behavior that the new version changes, or the move to the new
stack. Most verifiers also re-test behavior established in earlier steps, so
that a step generally cannot be passed at the expense of earlier work. The step
reward is binary: $r_k=1$ if the step passes every check, and $r_k=0$
otherwise. Optimization steps use a different formulation.

\paragraph{Optimization reward.}
An optimization step is graded on both correctness and speed, and its reward
is the product of two binary terms,
\begin{equation}
  \label{eq:opt-score}
  r_k = c \cdot v,
\end{equation}
where $c=1$ if the library's outputs pass the correctness gate (which covers
both the functions of step $k$ and those optimized in earlier steps) and $c=0$
otherwise, and $v=1$ if the speedup $s$ exceeds the threshold $s_{\mathrm{ref}}$ and $v=0$
otherwise. Here $s$ is the geometric mean of the per-function speedups (each
floored at $1\times$), and $s_{\mathrm{ref}}$ is the speedup $s$ measured for the
step's reference solution, fixed in the step's verifier. \Cref{app:speedups}
reports the thresholds and the raw speedups.

\paragraph{Solving a task.}
A rollout \emph{solves} its task if it passes every step ($r_k=1$ for all
$k$). A solution must therefore complete every milestone of a growing system
rather than produce only a final artifact.

\subsection{Preventing Reward Hacking}
\label{sec:reward-hacking}

Reward hacking, in which an agent obtains reward without accomplishing the
intended task \citep{amodei2016concrete,skalse2022defining}, has been observed
in frontier models on coding and optimization tasks
\citep{baker2025monitoring,metr2025rewardhacking}. Because each verifier in
\bench builds and runs code written by the agent, we design our verifiers to
treat this code as potentially adversarial and to block four kinds of shortcut
(\Cref{tab:hacks}).

\begin{table}[t]
  \centering
  \footnotesize
  \begin{tabularx}{\linewidth}{@{}>{\raggedright\arraybackslash}p{2.2cm} >{\raggedright\arraybackslash}X >{\raggedright\arraybackslash}p{2.6cm} >{\raggedright\arraybackslash}X@{}}
    \toprule
    Task & Attack & Result before the fix & Countermeasure \\
    \midrule
    SMB server & A background process rewrites the reward file with a reward
      of 1
      & step passed
      & the score is passed through a channel the agent cannot write \\
    Arquillian & A planted \texttt{JUnitCore} class reports every test as
      passing
      & step passed
      & the scorer's own test runner comes first on the classpath \\
    Go toolchain & A C program reads the expected outputs by absolute path
      & 62 of 133 cases pass
      & graded code runs unprivileged; outputs are readable only by root \\
    Protocol Buffers & A C wrapper runs the official \texttt{protoc}, shipped
      as a compressed file and unpacked during the build
      & step passed
      & every file the agent can write is scanned for upstream code, including
        inside compressed files \\
    CityVoice (Rails) & Deprecation warnings are turned off through
      configuration
      & deprecation check passes
      & warnings are captured inside the running process \\
    BoringSSL & A cache returns stored outputs for repeated inputs
      & $194\times$ speedup on an AES function (reference solution: $38\times$)
      & inputs are perturbed on every timed call \\
    \bottomrule
  \end{tabularx}
  \caption{Examples of attacks, found during adversarial hardening, that
    succeeded against early versions of the verifiers, and the checks that now
    block them.}
  \label{tab:hacks}
\end{table}

\paragraph{Tampering with the grader.}
The verifier rebuilds the artifact from the agent's sources rather than using
any binary left in the workspace. It builds and runs the agent's code as an
unprivileged user, which can neither write the reward file nor read the
expected outputs, and terminates the agent's processes before computing the
reward. Optimization verifiers build the reference library and the timing
harness themselves and confirm that the timed function is the one the library
actually calls, so that the agent cannot alter what is measured. A step also
cannot be passed by
keeping tests from running: if the build fails, a test group runs fewer cases
than it declares, or grading exceeds its time limit, the step receives a reward
of zero.

\paragraph{Memorizing expected outputs.}
Beyond hiding the graded cases, optimization verifiers
generate their inputs at grading time and check
correctness and speed on different inputs, so a function cannot return outputs
that it stored during the correctness check. Official conformance suites are
public, so an agent could hard-code their expected outputs; version upgrade
tasks therefore also grade modified versions of these tests, such as Ethereum
fixtures with changed gas limits and balances, whose expected outputs the agent
has never seen.

\paragraph{Delegating the work.}
In the 56 tasks without public network access, the agent can reach only the
model API, and optimization tasks remove the scored kernels together with their
variants for other processor architectures. The verifiers scan the workspace
for foreign executables and runtimes and monitor graded programs for hand-offs
to other processes; the PHP and Ruby verifiers also compare the token structure
of the agent's sources with upstream's to detect copies with renamed
identifiers.

\paragraph{Passing checks without doing the work.}
We design each check so that only a working implementation passes it. The CSS
layout verifier credits an element only if all of its layout values are
correct, so a program that always outputs \texttt{display: block}, the correct
value for most elements, does not pass. The Lua verifier includes programs that
must stop with an error, so an interpreter whose \texttt{assert} never fails
cannot pass by running every program to completion. Each step's new requirements are
tested with cases that the previous step's code fails, so leaving that code
unchanged earns nothing. Migration verifiers query the running system rather than its configuration
files, for example reading the Rails version from the booted application rather
than from the Gemfile. They also capture deprecation warnings inside the running
process, where configuration cannot turn them off, so deprecated APIs must be
removed rather than hidden.

\paragraph{Adversarial hardening.}
During development, we attacked the verifiers with submissions designed to earn
reward without doing the work. Each successful attack led to a new check, which
we tested against the attack and against the reference solutions.

\subsection{Benchmark Statistics}
\label{sec:stats}

\Cref{tab:stats} summarizes the benchmark.

\begin{table}[h!]
  \centering
  \small
  \setlength{\tabcolsep}{6pt}
  \renewcommand{\arraystretch}{1.18}
  \newcommand{\grp}[1]{\multicolumn{2}{@{\hspace{4pt}}>{\columncolor{black!6}[4pt][4pt]}p{\dimexpr\linewidth-8pt\relax}@{\hspace{4pt}}}{\scriptsize\bfseries\MakeUppercase{#1}}\\}
  \begin{tabularx}{\linewidth}{@{\hspace{4pt}}>{\raggedright\arraybackslash}p{3.6cm} X@{\hspace{4pt}}}
    \toprule
    \grp{Scale}
    Tasks & \textbf{69}: optimization (18) $\cdot$ feature implementation (18) $\cdot$
      version upgrade (20) $\cdot$ migration (13) \\
    Steps & \textbf{574} in total; \textbf{5--15} per task (median 8, mean 8.3) \\
    \grp{Budget}
    Agent budget per step & 2\,h (471 steps) $\cdot$ 3\,h (56 steps) $\cdot$ 4\,h (47 steps) \\
    Agent budget per task & 10--60\,h (median 16\,h) \\
    \grp{Setting}
    Starting point & empty or skeleton workspace (34 tasks) $\cdot$ existing
      open-source codebase (35 tasks) \\
    Languages (tasks) & C/C++ (38) $\cdot$ Rust (11) $\cdot$ Java (6)
      $\cdot$ Go (4) $\cdot$ JavaScript (4) $\cdot$ Python (3) $\cdot$ PHP (2)
      $\cdot$ Ruby (1) \\
    Network during agent work & model API only (56 tasks) $\cdot$ public
      access (13 migration tasks) \\
    \bottomrule
  \end{tabularx}
  \caption{\bench at a glance.}
  \label{tab:stats}
\end{table}

Many steps constitute substantial pieces of work in their own right. In
from-scratch tasks, the first step asks for a complete working system, and its
reference solution is correspondingly large: about 37,000 lines of C for the
PHP~5.6 interpreter, 42,000 lines of Java for the Redis~2.8-compatible server,
18,000 lines of C for the CRuby~1.8.7-compatible interpreter, and 11,000 lines
of C for the Kubernetes~1.25 control plane. Later steps change less code but
still require long sessions: in our experiments, agents spent a median of 45
minutes on each of them.

The tasks cover a wide range of software:
language implementations (interpreters for Lua, PHP, and Ruby; a Go toolchain;
a Java virtual machine; a Protocol Buffers compiler);
databases and storage (SQLite-compatible engines, PostgreSQL, Redis, Apache Iceberg,
Icechunk);
networking (QUIC and HTTP/3, TLS, SMB, SMTP and IMAP, Matrix, Tor, SRT);
cryptography (BoringSSL, ML-KEM, ML-DSA, Certificate Transparency);
blockchains (the Ethereum Virtual Machine, a Zcash node);
media and documents (H.264, VVC, JPEG XS, HTJ2K, FFmpeg, PDF, OpenType, CSS layout);
systems tools (an ELF linker, a DWARF toolkit, a Kubernetes control plane,
git, the uv Python package manager);
scientific computing (LAMMPS, sequence alignment, pandas, Spark, point clouds);
healthcare (a FHIR server);
and web applications (Rails, React, Java EE).
\Cref{app:tasks} lists all tasks.

% ===========================================================================
\section{Experiments}
\label{sec:experiment}

\subsection{Setup}
\label{sec:rollout-setup}

We evaluate nine frontier models: Claude Opus 5.5, Claude Fable 5.1, Gemini 4
Argon, GPT-6.1 Sol, GPT-6 Astra, DeepSeek V4.1 Flash, Qwen 3.8 Max, MiMo 2.6
Pro, and Grok 4.7. Each model runs in an agent harness provided by Harbor
(Claude Code, Codex CLI, mini-swe-agent, or Grok Build) at the highest
reasoning effort that the harness supports (\Cref{tab:harness}); the results
therefore reflect each model in combination with its harness. Each model is run
four times per task.

\paragraph{Metrics.}
We report the \textbf{solve rate}, defined as the fraction of rollouts that
solve their task by receiving full credit at every step, and \textbf{Pass@4}, defined as the fraction of
tasks solved by at least one of the four rollouts. To measure progress short of
a complete solution, we also report the fraction of \textbf{steps passed} with
full credit.

\subsection{Main Results}
\label{sec:main-results}

\Cref{tab:results} summarizes the results. We highlight three findings.

\begin{table}[t]
  \centering
  \small
  \setlength{\tabcolsep}{4pt}
  \setlength{\aboverulesep}{0pt}
  \setlength{\belowrulesep}{0pt}
  \renewcommand{\arraystretch}{1.22}
  \begin{tabular}{l >{\columncolor{solvedbg}}c c c c c c @{\hspace{14pt}} c}
    \toprule
    & \multicolumn{6}{c}{\textbf{Tasks solved}} & \\
    \cmidrule(lr){2-7}
    Model & \textbf{Overall} & Optimization & Feature & Upgrade & Migration & Pass@4 & Steps passed \\
    \midrule
    Claude Opus 5.5     & \textbf{18.5} \ci{11.6}{25.7} & 34.7 & \textbf{12.5} & \phantom{0}\textbf{8.8} & \textbf{19.2} & \textbf{31.9} & \textbf{56.8} \\
    Claude Fable 5.1    & 17.8 \ci{\phantom{0}9.8}{26.1} & \textbf{45.8} & \phantom{0}8.3 & \phantom{0}1.3 & 17.3 & 24.6 & 49.1 \\
    Gemini 4 Argon      & 10.1 \ci{\phantom{0}5.8}{15.2} & 16.7 & \phantom{0}6.9 & \phantom{0}3.8 & 15.4 & 27.5 & 40.1 \\
    GPT-6.1 Sol         & 10.1 \ci{\phantom{0}5.1}{15.9} & 12.5 & \phantom{0}6.9 & \phantom{0}5.0 & \textbf{19.2} & 21.7 & 37.7 \\
    GPT-6 Astra         & \phantom{0}6.9 \ci{\phantom{0}2.9}{12.0} & \phantom{0}5.6 & \phantom{0}5.6 & \phantom{0}2.5 & 17.3 & 15.9 & 36.7 \\
    DeepSeek V4.1 Flash & \phantom{0}6.5 \ci{\phantom{0}2.9}{10.9} & 11.1 & \phantom{0}2.8 & \phantom{0}1.3 & 13.5 & 15.9 & 35.8 \\
    Qwen 3.8 Max        & \phantom{0}4.0 \ci{\phantom{0}1.1}{\phantom{0}7.6} & \phantom{0}4.2 & \phantom{0}1.4 & \phantom{0}1.3 & 11.5 & \phantom{0}8.7 & 24.4 \\
    MiMo 2.6 Pro        & \phantom{0}2.9 \ci{\phantom{0}0.4}{\phantom{0}6.5} & \phantom{0}4.2 & \phantom{0}1.4 & \phantom{0}1.3 & \phantom{0}5.8 & \phantom{0}5.8 & 22.1 \\
    Grok 4.7            & \phantom{0}1.4 \ci{\phantom{0}0.0}{\phantom{0}4.0} & \phantom{0}0.0 & \phantom{0}0.0 & \phantom{0}1.3 & \phantom{0}5.8 & \phantom{0}2.9 & 13.9 \\
    \bottomrule
  \end{tabular}
  \caption{Results for nine models, in percent, with a 95\%
    bootstrap confidence interval over tasks for the overall solve rate. The
    family columns cover optimization (18 tasks), feature implementation (18),
    version upgrade (20), and migration (13). The best value in each column
    is shown in bold. Gemini 4 Argon and GPT-6.1 Sol tie on the overall solve
    rate and are ordered by Pass@4.}
  \label{tab:results}
\end{table}

\begin{figure}[t]
  \centering
  \begin{tikzpicture}
    \begin{axis}[
        xbar,
        width=0.78\linewidth,
        height=6.6cm,
        xmin=0, xmax=60,
        xtick={0,10,20,30,40,50,60},
        xlabel={Percent},
        symbolic y coords={Grok 4.7,MiMo 2.6 Pro,Qwen 3.8 Max,DeepSeek V4.1 Flash,GPT-6 Astra,GPT-6.1 Sol,Gemini 4 Argon,Claude Fable 5.1,Claude Opus 5.5},
        ytick=data,
        enlarge y limits=0.07,
        tick label style={font=\footnotesize},
        label style={font=\footnotesize},
        xmajorgrids,
        grid style={black!10},
        axis line style={black!35},
        axis x line*=bottom,
        axis y line*=left,
        ytick style={draw=none},
        xtick style={black!35},
        legend style={at={(0.5,1.02)}, anchor=south, legend columns=2,
                      draw=none, font=\footnotesize,
                      /tikz/every even column/.append style={column sep=10pt}},
        reverse legend,
        legend image code/.code={\draw[#1, draw=none] (0cm,-0.1cm) rectangle (0.4cm,0.1cm);},
      ]
      \addplot[fill=stepsfill, draw=none, bar width=11pt, bar shift=0pt]
        coordinates {(13.9,Grok 4.7) (22.1,MiMo 2.6 Pro) (24.4,Qwen 3.8 Max)
                     (35.8,DeepSeek V4.1 Flash) (36.7,GPT-6 Astra) (37.7,GPT-6.1 Sol)
                     (40.1,Gemini 4 Argon) (49.1,Claude Fable 5.1) (56.8,Claude Opus 5.5)};
      \addplot[fill=seriesB, draw=none, bar width=5pt, bar shift=0pt,
               nodes near coords={\pgfmathprintnumber[assume math mode=true, fixed,
                 fixed zerofill, precision=1]{\pgfplotspointmeta}},
               nodes near coords align={horizontal},
               every node near coord/.append style={font=\footnotesize\bfseries, text=black!85}]
        coordinates {(1.4,Grok 4.7) (2.9,MiMo 2.6 Pro) (4.0,Qwen 3.8 Max)
                     (6.5,DeepSeek V4.1 Flash) (6.9,GPT-6 Astra) (10.1,GPT-6.1 Sol)
                     (10.1,Gemini 4 Argon) (17.8,Claude Fable 5.1) (18.5,Claude Opus 5.5)};
      \legend{Steps passed (\%), Solve rate (\%)}
    \end{axis}
  \end{tikzpicture}
  \caption{Solve rate (orange, labeled) for each model, overlaid on the
    fraction of steps passed with full credit (light blue).}
  \label{fig:compound}
\end{figure}
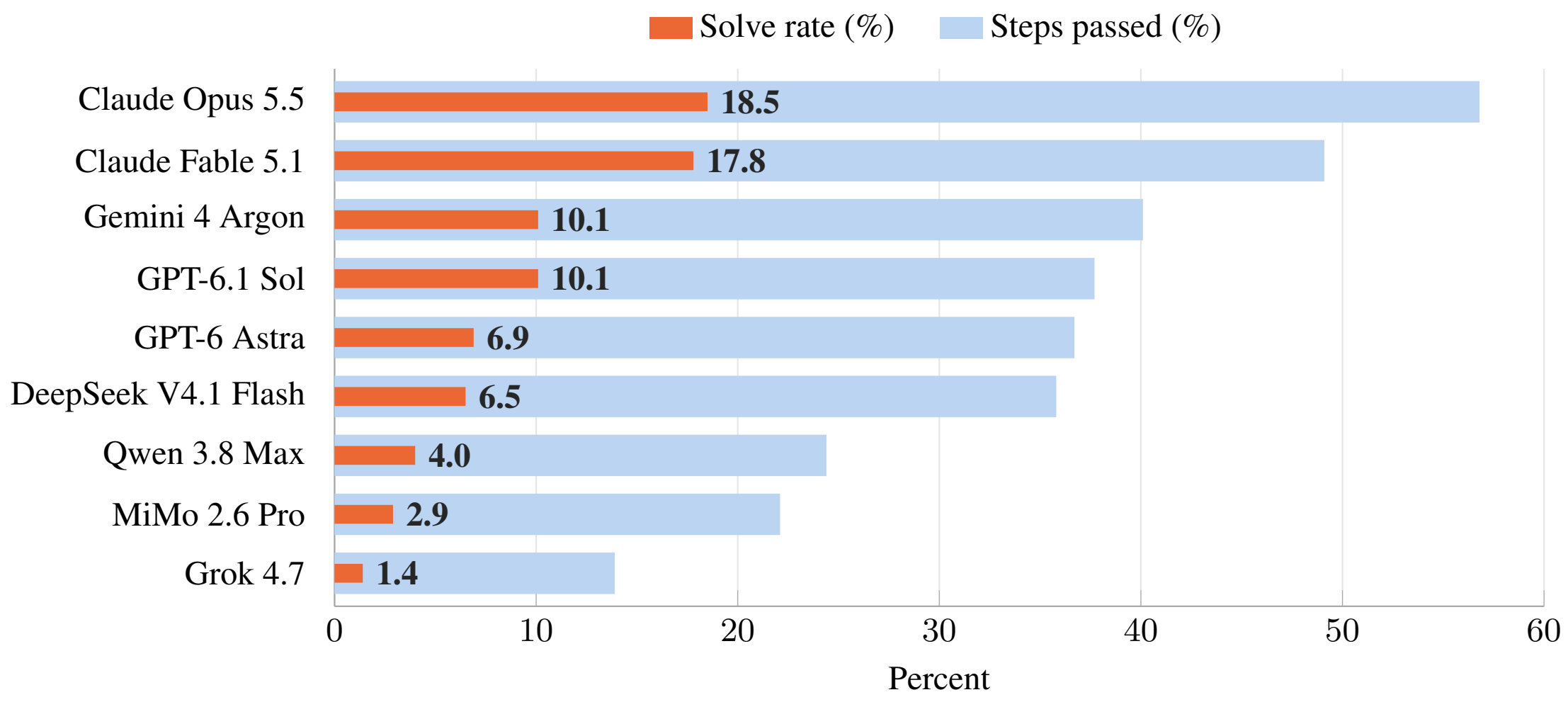

\paragraph{Whole tasks are seldom solved.}
The strongest model, Claude Opus 5.5, reaches a solve rate of 18.5\% and
solves 31.9\% of the tasks within four attempts, although it passes 56.8\% of
its individual steps (\Cref{fig:compound}). Every other model has a solve
rate of at most 17.8\%, and 38 of the 69 tasks are not solved by any model,
including 16 of the 20 version-upgrade tasks. Success is also inconsistent
across attempts at the same task: Gemini 4 Argon has a solve rate of 10.1\% but
solves 27.5\% of the tasks at least once, and Claude Fable 5.1 has a solve rate
of 17.8\% but solves 24.6\% of the tasks at least once.

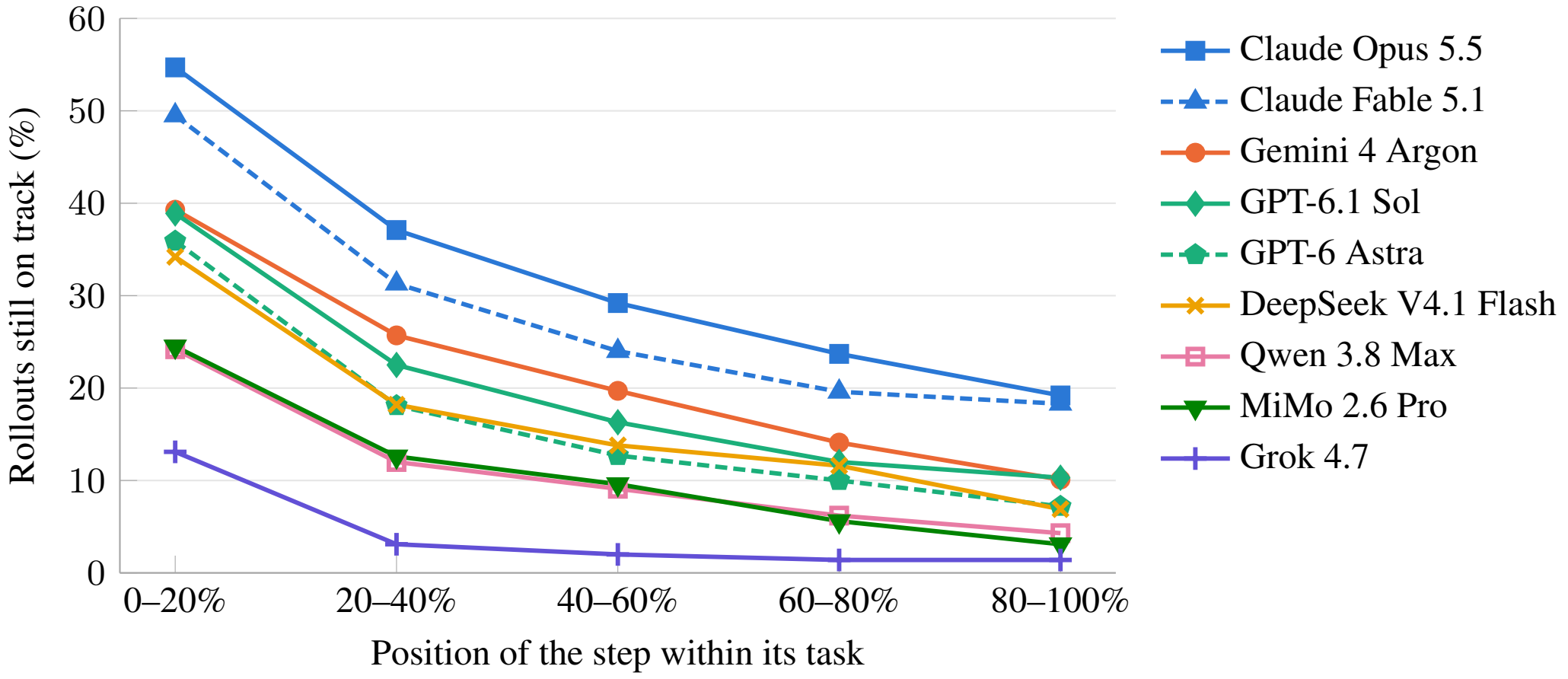
\begin{figure}[t]
  \centering
  \begin{tikzpicture}
    \begin{axis}[
        width=0.62\linewidth,
        height=6.4cm,
        xmin=0.75, xmax=5.25,
        ymin=0, ymax=60,
        ytick={0,10,20,30,40,50,60},
        xtick={1,2,3,4,5},
        xticklabels={0--20\%,20--40\%,40--60\%,60--80\%,80--100\%},
        xlabel={Position of the step within its task},
        ylabel={Rollouts still on track (\%)},
        tick label style={font=\footnotesize},
        label style={font=\footnotesize},
        ymajorgrids,
        grid style={black!10},
        axis line style={black!35},
        axis x line*=bottom,
        axis y line*=left,
        tick style={black!35},
        every axis plot/.append style={line width=1.1pt, mark size=1.9pt,
                                       mark options={solid}},
        legend cell align=left,
        legend style={at={(1.03,1)}, anchor=north west, draw=none,
                      font=\footnotesize, row sep=1pt},
      ]
      % Models in the order of the results table (by solve rate); one hue per provider, one marker per model.
      \addplot[color=seriesA, mark=square*]
        coordinates {(1,54.7) (2,37.1) (3,29.2) (4,23.7) (5,19.2)};
      \addplot[color=seriesA, dash pattern=on 3pt off 1.5pt, mark=triangle*, mark size=2.3pt]
        coordinates {(1,49.5) (2,31.3) (3,24.0) (4,19.6) (5,18.3)};
      \addplot[color=seriesB, mark=*]
        coordinates {(1,39.3) (2,25.7) (3,19.7) (4,14.1) (5,10.1)};
      \addplot[color=seriesC, mark=diamond*, mark size=2.4pt]
        coordinates {(1,38.9) (2,22.5) (3,16.3) (4,12.0) (5,10.3)};
      \addplot[color=seriesC, dash pattern=on 3pt off 1.5pt, mark=pentagon*, mark size=2.2pt]
        coordinates {(1,35.9) (2,18.1) (3,12.7) (4,10.0) (5,7.2)};
      \addplot[color=seriesD, mark=x, mark size=2.6pt]
        coordinates {(1,34.2) (2,18.2) (3,13.8) (4,11.6) (5,6.9)};
      \addplot[color=seriesE, mark=square, mark options={solid, fill=white}]
        coordinates {(1,24.2) (2,12.0) (3,9.1) (4,6.2) (5,4.3)};
      \addplot[color=seriesF, mark=triangle*, mark size=2.3pt, mark options={solid, rotate=180}]
        coordinates {(1,24.5) (2,12.6) (3,9.6) (4,5.6) (5,3.1)};
      \addplot[color=seriesG, mark=+, mark size=2.8pt]
        coordinates {(1,13.1) (2,3.1) (3,2.0) (4,1.4) (5,1.4)};
      \legend{Claude Opus 5.5, Claude Fable 5.1, Gemini 4 Argon, GPT-6.1 Sol,
              GPT-6 Astra, DeepSeek V4.1 Flash, Qwen 3.8 Max, MiMo 2.6 Pro,
              Grok 4.7}
    \end{axis}
  \end{tikzpicture}
  \caption{Fraction of rollouts that are still on track, by position of the
    step within the task, for each model over all 69 tasks. Each step is
    assigned to one of five equal ranges of relative position within its task. A rollout is on
    track at a step if it has passed that step and every earlier step. Values
    are averaged within each task and then across tasks.}
  \label{fig:position}
\end{figure}

\paragraph{Success declines along the task, and failures begin early.}
Over the course of a task (\Cref{fig:position}), the fraction of rollouts that
remain on track, having passed every step so far, falls from 54.7\% in the
first fifth of the steps to 19.2\% in the last fifth for Claude Opus 5.5, and
from 13.1\% to 1.4\% for Grok 4.7. For every model, the sharpest drop occurs
between the first and second fifths. Over all models, this fraction falls from
34.9\% to 9.0\%, whereas the fraction of steps passed declines more gradually,
from 40.7\% to 27.8\%. Step outcomes also depend strongly on the
outcome of the previous step: a step that follows a passed step is passed
70.2\% of the time, compared with 14.1\% for a step that follows a failed step.

\paragraph{Task families differ, and version upgrade is the hardest.}
Pooled over all models, the solve rate is 15.0\% for optimization, compared
with 13.9\% for migration, 5.1\% for feature implementation, and 2.9\% for
version upgrade. In version-upgrade tasks, the agent builds a system and then
changes its behavior release by release; no model has a solve rate above 8.8\%
on these tasks. In optimization tasks, a solution requires correct output and a
speedup above that of the reference solution at every step; Claude Fable 5.1
leads this family with a solve rate of 45.8\%, followed by Claude Opus 5.5 with
34.7\%, and 4 of the 18 optimization tasks are not solved by any model.

\subsection{Failure Analysis}
\label{sec:result-analysis}

We examine how failing sessions end and why the work fails.

\paragraph{Failing agents stop early and report success.}
Running out of time accounts for few failures. Only 12\% of failed steps run to within
five minutes of the time budget. For six of these models, the median failed
step uses 24--37\% of its budget, leaving 77--93 minutes unused; only Qwen 3.8
Max and MiMo 2.6 Pro often exhaust the budget (in 39\% and 24\% of their failed
steps). Agents' final reports show why they stop (we label a stratified sample
of 2,157 final reports; \Cref{app:details}). In 52\% of
failed steps, the agent reports the step as complete and names no failing check or missing
requirement. In another 26\%, it reports specific failing checks or missing
features and stops anyway, with a median of 66 minutes left. Only 3\% cite the
time budget as a reason for stopping, and 19\% end without a final report. A report of completion is therefore a
poor signal of success: agents report completion and name no shortcoming in 73\% of
passed steps, and when an agent does so, the verifier still fails the step 57\% of
the time, compared with 65\% of all steps. This rate ranges from 26\% and 38\% for
Claude Opus 5.5 and Claude Fable 5.1 to 82\% for Grok 4.7, whose steps fail 43\%,
51\%, and 86\% of the time overall. The two Claude models report gaps in 68\% and 52\% of their failed
steps, more often than the other models (at most 37\%), but they also stop with
time left.

\paragraph{Session length does not separate passing from failing work.}
Within the same step and model, sessions that pass and sessions that fail last
about equally long: across step--model pairs with both outcomes, the median
ratio of their durations is 1.05. Together with the early stops, this indicates
that agents are limited by when they judge the work to be complete rather than
by the clock. Whether a larger budget would help cannot be determined from
these runs, because failing agents seldom use the budget they already have.

\paragraph{Agents check their work against their own expectations.}
Agents typically stop once the public tests or their own tests pass. In one
Claude Fable 5.1 rollout on the
CSS layout task, the engine passed every public test but sized floats
incorrectly; the public tests contain 2 float-layout tests, and the hidden
tests contain 22. Tests that an agent writes encode its own reading of the
specification, so they confirm a misreading instead of exposing it. The passing
runs of the same steps differ mainly in what they check against: an
independent oracle or a differential comparison rather than the agent's own
expectations. On the TLS task, a passing Claude Opus 5.5 run built the BoGo
conformance runner from the local package registry and fixed the bugs that it
exposed, whereas a failing Gemini 4 Argon run tested only against OpenSSL. On
the DWARF task, a passing run compared the DWARF~4 and DWARF~5 builds of the
same programs, a comparison that exposes the range-list error that a failing
Claude Fable 5.1 run left in place. In a migration task, a GPT-6 Astra run
upgraded Guice from 2.0 to 5.1 without being asked, which changed the routing
of two HTTP requests, whereas a passing run that made the same upgrade compared
its behavior with the original application, noticed the difference, and
reverted the upgrade.

\paragraph{Agents explain away evidence of defects.}
In many failures, the agent's own session contained evidence of the defect,
but the agent explained it away or stopped regardless. In a DeepSeek V4.1 Flash rollout on the PDAL task, the agent
observed NaN where the documentation specifies 0, accepted it because it
matched the code inherited from the previous release, and stopped with more
than three hours left. In a Qwen 3.8 Max rollout on the mail-stack task, the
agent reversed the argument order of the IMAP \texttt{MOVE} command, contrary to
RFC~6851 in the workspace, to match a test that it had written itself. In a
Claude Fable 5.1 rollout on the Python~2 to 3.14 migration, the agent's own
type checker flagged 11 methods, which the agent judged acceptable.

% ===========================================================================
\section{Related Work}
\label{sec:related}

\paragraph{Sequential and evolving codebases.}
The most closely related work evaluates agents on sequences of dependent
changes and consistently reports that performance degrades as work
accumulates. SlopCodeBench \citep{orlanski2026slopcodebench} contains 36 Python
problems with 196 checkpoints built from an empty workspace; each checkpoint
continues from the workspace of the previous checkpoint, and its strict score
reruns all earlier tests. SWE-Chain \citep{lam2026swechain} chains 155
release-level upgrades of Python packages and counts tests that pass before a
step and fail after it as regressions. EvoCode-Bench
\citep{shen2026evocodebench} contains 26 tasks of 5--15 rounds, each of which
re-verifies the requirements of earlier rounds that remain valid.
SWE-Milestone \citep{deng2026swemilestone} orders 98 milestones mined from
seven repositories in five languages, and SWE-STEPS \citep{shastry2026swesteps}
evaluates sequences of pull requests in Python repositories. Like these
benchmarks, \bench carries the agent's own workspace forward from step to step.
\bench differs in breadth, in grounding, and in the scale of each step, which
is a multi-hour engineering request. Its tasks span nine languages and four
families of engineering work, including performance optimization and stack
migration; they include both systems built from scratch and existing
codebases; and each step follows a public progression and is graded, wherever
possible, by the field's own conformance suites and reference tools.

\paragraph{Issue-level software engineering.}
SWE-bench \citep{jimenez2023swebench} asks a model to resolve a real GitHub
issue and scores the final patch with the issue's tests, and SWE-agent
\citep{yang2024sweagent} studies agent--computer interfaces in this setting.
SWE-bench Verified \citep{openai2024swebenchverified} retains the subset of
SWE-bench tasks that human annotators judged to have clear problem statements
and fair tests. \bench shares their use of real repositories and executable
tests, but replaces the single change with a chain of dependent changes and
derives both its instructions and its verifiers from existing sources of truth
rather than from issue text and the tests of a single fix.

\paragraph{Multilingual and migration benchmarks.}
Multi-SWE-bench \citep{zan2025multiswebench} extends issue resolution to 1,632
instances in seven languages other than Python, and SWE-Bench Pro
\citep{deng2025swebenchpro} collects 1,865 long-horizon tasks from 41
repositories, including private commercial ones. MigrationBench
\citep{liu2025migrationbench} evaluates repository-level migration of Java~8
projects to Java~17 and 21. These benchmarks broaden language coverage or
target one kind of work, but each of their tasks is a single change. \bench
covers nine languages and treats migration as one of four families, carried out
step by step; for example, one task moves a Java project from Java~6 to
Java~25 in 13 steps.

\paragraph{Long-horizon agentic tasks.}
Terminal-Bench \citep{merrill2026terminalbench} contains 89 difficult tasks in
command-line environments; each trial is a single attempt at one task, graded
on the final container state. DeepSWE \citep{huang2026deepswe} contains 113
feature requests against existing open-source repositories, each attempted in
a single rollout and graded as pass or fail by a functional verifier that also
checks that the repository's existing behavior is preserved. FrontierSWE
\citep{chu2026frontierswe} contains 17 very long tasks that allow up to 20
hours per run and are graded with continuous partial credit. TheAgentCompany
\citep{xu2024theagentcompany} evaluates agents on workplace tasks in a
simulated software company. \bench shares their focus on long, realistic work
but differs in structure: the work is a chain of long steps, each step is
verified separately, and each step continues from the code that the agent
itself wrote.

\paragraph{Building software from scratch.}
Commit0 \citep{zhao2024commit0} asks agents to generate a library from scratch
against its unit tests, and ProgramBench \citep{yang2026programbench} asks
agents to rebuild a program in any language from only its compiled executable
and documentation, graded by hidden behavioral tests. The from-scratch tasks of
\bench do not end with the first working version: after building a system, the
agent carries it through later versions of its specification, as real projects
evolve.

\paragraph{Performance optimization.}
SWE-Perf \citep{he2025sweperf} and SWE-fficiency \citep{ma2025swefficiency}
evaluate repository-level optimization against expert speedups. The
optimization family of \bench complements them with sequences of kernel
families in a single workspace and with low-level targets such as SIMD and
assembly kernels in cryptography, media, and numerical libraries.

\paragraph{Code generation.}
HumanEval \citep{chen2021humaneval}, LiveCodeBench
\citep{jain2024livecodebench}, and RepoBench \citep{liu2023repobench} evaluate
function-level generation, contamination-free generation, and repository-level
completion, respectively.

In summary, \bench combines the step-by-step protocol of sequential benchmarks
with the long, realistic tasks of single-episode benchmarks, and adds breadth
(nine languages, four families of work, and both from-scratch and existing
codebases) and grounding (every step is tied to a public progression and
graded, wherever possible, by the field's own tools).

% ===========================================================================
\section{Limitations and Future Work}
\label{sec:future-work}

Despite expert review and grounding in existing sources of truth, some tasks may
still contain an ambiguous requirement or a verifier that misses or wrongly
rejects a behavior; we release the benchmark so that the community can report
and fix such problems. Language coverage is uneven, with C and C++ used in 38 of
the 69 tasks. Because the tasks derive from public projects, memorization
cannot be ruled out, although most tasks block network access
(\Cref{app:details}). The full-credit threshold for optimization steps is the
speedup of our reference solution, which need not be the best speedup
achievable.

Future work includes an isolated-step baseline, in which step $k$ starts from
the reference solution of steps $1,\dots,k-1$, to separate accumulated errors
from step difficulty
\citep{shen2026evocodebench,deng2026swemilestone,shastry2026swesteps}, and an
extension to multimodal, UI-centric work such as web front-end development.

% ===========================================================================
\section{Conclusion}
\label{sec:conclusion}

\bench evaluates coding agents in the way developers use them: as a sequence of
multi-hour requests, each of which builds on the agent's own code and is
verified separately. Its tasks follow real progressions (language releases,
protocol revisions, upstream release histories, framework migrations, and
hand-optimized kernels), so every step has a well-defined target, and the
instruction and the verifier are derived from the same source of truth. Whole
tasks are seldom solved: the strongest model reaches a solve rate of only
18.5\%, and
more than half of the tasks are not solved by any model. We hope that \bench
will help measure progress toward agents that can carry a codebase through long
sequences of real engineering work. We welcome contributions from the
community to further improve the benchmark.

\bibliographystyle{unsrtnat}
\bibliography{references}

% ---------------------------------------------------------------------------
% arXiv submission note:
% arXiv does not run BibTeX. Before uploading, run BibTeX locally and paste the
% contents of main.bbl into a thebibliography environment below, replacing the
% \bibliographystyle and \bibliography lines above. See README.md.
% ---------------------------------------------------------------------------

\clearpage
\appendix
{\centering\LARGE\bfseries Appendix\par}
\vspace{1.2em}

% ===========================================================================
\section{Task Format and Protocol}
\label{sec:protocol}

Every task is packaged in the multi-step task format of Harbor
\citep{harbor2026}. \Cref{fig:layout} shows the layout of one task. The root
\texttt{task.toml} declares the ordered steps and, for each step, the agent's
time budget and the verifier's time limit; it also specifies the container's
resources, the agent's network policy, and the reward strategy. The
\texttt{environment/} directory contains the Dockerfile that builds the
starting workspace (an empty directory, a skeleton, or a repository pinned at
an upstream commit) together with every toolchain that the task requires. Each
step has its own directory with up to four components:
\begin{itemize}
  \item \texttt{instruction.md}, the instruction that the agent receives. Most
    instructions follow a common template: what the step must newly achieve,
    including the interface that the verifier will use; the specifications and
    documents in the workspace that define the target; and the rules of the
    task, such as the time budget and the prohibition on copying an existing
    implementation. Some tasks number every requirement and grade exactly that
    list.
  \item \texttt{workdir/}, files that Harbor copies into the workspace before
    the step starts, typically the specification of the next release, an
    updated test-interface contract that describes how outputs are compared,
    and new public tests (535 steps add files in this way).
  \item \texttt{tests/}, the step's verifier: a Dockerfile that builds a
    separate verifier image, and a \texttt{test.sh} script that writes the
    step's reward to \texttt{reward.json}. Verifiers share a common structure:
    they isolate the grading process from the agent's code, rebuild the
    artifact from the agent's sources, check the rules stated in the
    instruction, and score the new milestone.
  \item \texttt{solution/}, the reference solution (\texttt{solve.sh}).
\end{itemize}
Because the format is open and standardized, the tasks run under any agent
harness that Harbor supports, and a task can be extended with new steps without
modifying the existing ones.

\begin{figure}[H]
  \centering
  \begin{minipage}{0.95\linewidth}
  \footnotesize
\begin{verbatim}
lua-vm-version-chain/
|-- task.toml          ordered steps; per-step agent and verifier budgets;
|                      resources, network policy, reward strategy (mean)
|-- environment/
|   |-- Dockerfile     starting workspace and toolchain (JDK 21; no network)
|   `-- timer.sh       reports the remaining budget to the agent
`-- steps/
    |-- 01-lua-5-1/
    |   |-- instruction.md   the step's contract
    |   |-- workdir/         copied into the workspace before the step:
    |   |                    SPEC/lua-5.1/ (reference manual); tests/ (official
    |   |                    Lua 5.1 suite, sample conformance programs,
    |   |                    local runner); WORKSPACE_HANDOFF.md
    |   |-- tests/           the verifier: Dockerfile (separate image),
    |   |                    test.sh, conformance corpus, scoring scripts
    |   `-- solution/        reference solution: solve.sh and its sources
    |-- 02-lua-5-2/   ...
    `-- 05-lua-5-5/   ...
\end{verbatim}
  \end{minipage}
  \caption{Layout of a task in the Harbor multi-step format, using the Lua task
    as an example. Each step has its own instruction, workspace additions,
    verifier, and reference solution.}
  \label{fig:layout}
\end{figure}

Harbor runs the steps in order. Before step $k$, Harbor
copies the step's \texttt{workdir/} into the workspace; the agent then
receives instruction $k$ and works on the workspace left by step $k-1$. Each
step the agent works autonomously
over many tool calls, and the workspace is graded only after the agent stops.
Most tasks ask the agent to maintain a handoff note in the workspace. When the
agent finishes or its budget expires, Harbor copies the workspace into the
step's verifier container. Build outputs can be excluded from this copy (the
Lua task, for instance, excludes them), so that the verifier rebuilds the
artifact from the agent's sources and nothing the agent writes can influence
how the step is graded. Every step runs and is scored even if an earlier step
received a reward of zero, and the agent can repair earlier work in later
steps.

Each step has a wall-clock budget of 2, 3, or 4 hours, which a timer in the
workspace reports to the agent. In 56 of the 69 tasks, the agent cannot reach
any network host other than the model API, so an existing implementation cannot
be downloaded. The 13 migration tasks run with public network access.

\Cref{tab:examples} traces four steps from different tasks, from the
instruction to the verifier, and \Cref{app:evm} shows how one task evolves
across all of its steps.

\begin{table}[H]
  \centering
  \footnotesize
  \begin{tabularx}{\linewidth}{@{}>{\raggedright\arraybackslash}X >{\raggedright\arraybackslash}X@{}}
    \toprule
    Step and instruction & What the verifier checks \\
    \midrule
    \textbf{Lua interpreter in Java, Lua 5.3 $\to$ 5.4.} Implement Lua~5.4 as
      its reference manual specifies, keeping the build and command-line
      interface and matching the reference interpreter's output and errors.
    & Runs the official Lua~5.4 test suite and conformance programs recorded
      from the reference interpreter, including programs that a correct
      interpreter must reject. \\
    \addlinespace
    \textbf{turso database engine in Rust, a later release.} Add
      \texttt{ALTER TABLE} column operations, checkpoint modes, and the other
      numbered requirements of the next upstream release.
    & Runs the engine's shell and compares its output and database files with
      those of stock \texttt{sqlite3}. \\
    \addlinespace
    \textbf{simdutf in C++, UTF-8 decoding.} Make UTF-8 validation and
      transcoding faster without changing any output byte or reported error
      position.
    & Checks for identical output on hidden inputs against an unmodified
      reference implementation, then times each function on separate hidden
      inputs; the reward is 1 only if the output is exact and the speedup
      exceeds that of the reference solution, $s_{\mathrm{ref}}$. \\
    \addlinespace
    \textbf{CityVoice Rails application, Rails 4.0 $\to$ 4.2.} Move to
      Rails~4.2 and Ruby~2.3, retire deprecated APIs rather than silence them,
      and keep behavior unchanged apart from documented Rails changes.
    & Replays behavior recorded from the unmodified original (HTTP endpoints,
      webhooks, rake tasks, and mail) and checks that the migration is
      genuine, for example that the application actually boots on Rails~4.2. \\
    \bottomrule
  \end{tabularx}
  \caption{Four example steps, from instruction to verifier.}
  \label{tab:examples}
\end{table}

% ===========================================================================
\section{Example Task: The Ethereum EVM Task}
\label{app:evm}

The Ethereum EVM task asks the agent to implement go-ethereum's \texttt{evm}
tool in C, including its \texttt{t8n} (state transition), \texttt{t9n}
(transaction validation), and \texttt{b11r} (block building) subcommands, and
to carry the implementation through 14 Ethereum hard forks, from Frontier
to Osaka (\Cref{tab:evm}). The workspace initially contains the Yellow Paper,
go-ethereum's specification of the tool's interface with its test data, and a
script that runs public sample cases. Each step adds a sample of public cases
for its fork, and each subsequent step also adds the EIPs that define the next
fork. The agent works
without network access and has a budget of 2 hours per step.

The verifier of each step rebuilds the tool from source and replays a corpus
of public Ethereum test fixtures (GeneralStateTests, BlockchainTests,
execution-spec-tests, TransactionTests, and DifficultyTests), together with deterministic variations
of these fixtures, and compares state roots, receipts, post-states, transaction
validity, and block hashes with the expected results. The reference solution
comprises about 7,900 lines of C, plus a vendored BLS12-381 library for the KZG
and BLS12-381 precompiles.

Three rollouts solve the task, two by Claude Opus 5.5 and one by GPT-6.1 Sol,
each in 6.4 to 7.5 hours of agent time. Per-step success stays at 13 to 16
of the 36 rollouts from Frontier to Istanbul, falls to 10 to 12 from Berlin to
Shanghai, and drops to 3 to 6 in the last three forks (Cancun, Prague, and
Osaka), which introduce blob transactions, BLS12-381 precompiles, set-code
transactions, and the secp256r1 precompile.

\begin{table}[H]
  \centering
  \footnotesize
  \setlength{\tabcolsep}{4pt}
  \begin{tabularx}{\linewidth}{@{}r l >{\raggedright\arraybackslash}X r c@{}}
    \toprule
    Step & Fork & New in this fork & Cases & Full credit \\
    \midrule
    1 & Frontier & the full EVM: instruction set, gas schedule, tries, four precompiles & 1,166 & 15 \\
    2 & Homestead & \texttt{DELEGATECALL}; contract-creation and signature rules & 2,606 & 16 \\
    3 & Tangerine Whistle & repricing of I/O-heavy operations; 63/64 call-gas rule & 1,390 & 15 \\
    4 & Spurious Dragon & replay protection, state clearing, code-size limit & 1,044 & 13 \\
    5 & Byzantium & \texttt{REVERT}, \texttt{STATICCALL}, return data, alt\_bn128 and modexp precompiles & 2,493 & 14 \\
    6 & Constantinople & bitwise shifts, \texttt{CREATE2}, \texttt{EXTCODEHASH} & 3,122 & 14 \\
    7 & Istanbul & BLAKE2 precompile, \texttt{CHAINID}, \texttt{SELFBALANCE}, net gas metering & 2,334 & 15 \\
    8 & Berlin & typed transactions, access lists, cold and warm access costs & 2,484 & 12 \\
    9 & London & EIP-1559 base fee, \texttt{BASEFEE}, smaller refunds & 2,472 & 10 \\
    10 & Paris & proof of stake: \texttt{PREVRANDAO} replaces \texttt{DIFFICULTY} & 2,226 & 10 \\
    11 & Shanghai & \texttt{PUSH0}, initcode limit, withdrawals & 3,848 & 11 \\
    12 & Cancun & blob transactions, transient storage, \texttt{MCOPY} & 2,323 & 6 \\
    13 & Prague & BLS12-381 precompiles, set-code transactions, calldata repricing & 2,364 & 3 \\
    14 & Osaka & \texttt{CLZ}, secp256r1 precompile, modexp bounds, transaction gas cap & 3,639 & 6 \\
    \bottomrule
  \end{tabularx}
  \caption{The 14 steps of the Ethereum EVM task. Cases is the number of
    verifier cases for the new fork. Full credit is the number of the 36
    rollouts that receive full credit on the step.}
  \label{tab:evm}
\end{table}

% ===========================================================================
\section{Task List}
\label{app:tasks}

\Cref{tab:tasklist} lists all 69 tasks. ``Start'' is S for an empty or
skeleton workspace and E for an existing codebase.

{\footnotesize
\setlength{\tabcolsep}{4pt}
\begin{longtable}{@{}>{\raggedright\arraybackslash}p{3.8cm} >{\raggedright\arraybackslash}p{8.6cm} l c c@{}}
  \toprule
  Task & Progression across steps & Language & Steps & Start \\
  \midrule
  \endfirsthead
  \toprule
  Task & Progression across steps & Language & Steps & Start \\
  \midrule
  \endhead
  \bottomrule
  \endfoot
  \bottomrule
  \addlinespace[4pt]
  \caption{All 69 tasks in \bench.}\label{tab:tasklist}\\
  \endlastfoot
  \multicolumn{5}{@{}l}{\textit{Migration (13 tasks)}} \\
  Bento packaging & \texttt{setup.py} to PEP 517/621, uv lockfile, abi3 and free-threaded wheels, PEP 751 & Python & 6 & E \\
  Arquillian & Maven and Java 6 to Gradle and Java 25 & Java & 13 & E \\
  CityVoice & Rails 4.0 on Ruby 2.1 to Rails 8.1 on Ruby 3.4 & Ruby & 13 & E \\
  hedis (Redis fork) & IPv4-first networking to IPv6-only & C & 6 & E \\
  ISE server framework & C++98 to C++26 & C++ & 11 & E \\
  mini-git-server & Java EE 6 to Jakarta EE 11 and Spring Boot 4 & Java & 10 & E \\
  libStorage & GOPATH project to Go 1.25 & Go & 8 & E \\
  mmtf-spark & Spark 2.4 on Scala 2.11 to Spark 3.4 on Scala 2.13 and JDK 17 & Java & 5 & E \\
  WebGoatPHP & MySQL 5.7 to PostgreSQL 17, Latin-1 to UTF-8 & PHP & 6 & E \\
  TPM 2.0 simulator & vendored OpenSSL 1.0.2 to system OpenSSL 3.5 & C & 10 & E \\
  skutil & pandas 0.23 and NumPy 1.14 to pandas 3.0 and NumPy 2.3 & Python & 8 & E \\
  sublime-elasticsearch-client & Python 2 to Python 3.14 & Python & 7 & E \\
  React dashboard & React 15 and webpack to React 19, Vite, and the React Compiler & JavaScript & 8 & E \\
  \midrule
  \multicolumn{5}{@{}l}{\textit{Optimization (18 tasks)}} \\
  Apache Arrow & bit-unpacking, byte-stream split, Bloom filters, row kernels, swiss-join gather & C++ & 5 & E \\
  AviSynth+ & one video-filter family per step (overlay, blur, layer, turn, \ldots) & C++ & 7 & E \\
  BitMagic & one bit-block kernel family per step & C++ & 6 & E \\
  BoringSSL & AES, AES-GCM, GCM-SIV, Montgomery multiplication, P-256 & C++ & 5 & E \\
  FFmpeg libswscale & input conversion through YUV-to-RGB, in dataflow order & C & 6 & E \\
  ISA-L crypto & one multi-buffer hash per step (SHA-256, SHA-1, SHA-512, \ldots) & C & 7 & E \\
  LAMMPS & one Intel pair style per step (Lennard-Jones to AIREBO) & C++ & 5 & E \\
  mldsa-native (ML-DSA) & NTT through \texttt{polyz} unpacking & C & 6 & E \\
  mlkem-native (ML-KEM) & rejection sampling, NTT, reduction, (de)compression & C & 7 & E \\
  OpenJPH (HTJ2K) & block decoder and encoder, wavelets, color transform, packing & C++ & 6 & E \\
  parasail & local, global, and semi-global alignment: scores, statistics, traces & C & 7 & E \\
  Simd Library & one image-processing family per step (resizing, blur, edges, HOG, \ldots) & C++ & 8 & E \\
  simdutf & UTF-8, UTF-16, UTF-32, Latin-1, base64, validation & C++ & 7 & E \\
  srsRAN 5G PHY & LDPC encode, rate dematching, decode, precoding, demodulation, compression & C++ & 6 & E \\
  StringZilla & UTF-8 indexing, segmentation, case folding, hashing & C & 6 & E \\
  SVT-JPEG-XS & encoder and decoder pipeline, one stage per step & C & 8 & E \\
  uvg266 (VVC) & transforms, block cost, intra, motion compensation, quantization, loop filter & C & 6 & E \\
  Vortex & population count, bool packing, mask compress, filter, rank & Rust & 5 & E \\
  \midrule
  \multicolumn{5}{@{}l}{\textit{Feature implementation (18 tasks)}} \\
  Certificate Transparency & certificate-transparency-go v1.0 to v1.3.3 & Go & 13 & S \\
  ELF linker & static x86-64 to four architectures with LTO & Rust & 9 & S \\
  go-ethereum \texttt{evm t8n} & Berlin to Osaka & Go & 7 & E \\
  Apache Iceberg & format core to transaction actions & Rust & 5 & S \\
  Icechunk & spec 1 to spec 2.1 & Rust & 11 & S \\
  Tor client & circuits and a SOCKS proxy to bridges and pluggable transports & Rust & 12 & S \\
  OOXML spreadsheet engine & OPC writer to a formula and number-format engine & Go & 10 & S \\
  OpenSCENARIO engine & ten upstream releases of a scenario engine & C++ & 10 & E \\
  PCRE2 & 10.20 to 10.48 & C & 13 & S \\
  PDAL & 2.0.1 to 2.7.1 & C++ & 7 & E \\
  uv package manager & uv 0.1 to 0.12, from the \texttt{uv pip} interface to projects and managed Python & Rust & 15 & S \\
  QUIC and HTTP/3 & IETF draft-23 to RFC 9000 and later releases & Rust & 15 & S \\
  SQLite-compatible database & SQLite 3.8.0 to window functions & C & 5 & S \\
  SRT transport & UDT core to bonding and AEAD & C++ & 11 & S \\
  TLS 1.2/1.3 stack & rustls 0.19 to 0.23 & Rust & 13 & S \\
  tree-sitter & 0.19 and eight later releases & Rust & 9 & S \\
  turso & 15 releases, from a B-tree and WAL to views and encryption & Rust & 15 & S \\
  Zcash node & wire format to a JSON-RPC surface & Rust & 14 & E \\
  \midrule
  \multicolumn{5}{@{}l}{\textit{Version upgrade (20 tasks)}} \\
  CSS layout engine & CSS 2.1 to CSS Grid Level 1 & C & 5 & S \\
  DWARF toolkit & DWARF 2/3 to DWARF 5, then a debugger & Java & 5 & S \\
  Mail stack & RFC 821/822 SMTP to IMAP4rev2 and JMAP & C++ & 10 & S \\
  Ethereum EVM & Frontier to Osaka & C & 14 & S \\
  FHIR server & R4, then DSTU2/STU3 and R4B/R5, to the R6 ballot & C & 6 & S \\
  git & object model to SHA-256 and repository format v1 & PHP & 8 & S \\
  Go toolchain & Go 1.0 to Go 1.25 & C & 9 & S \\
  H.264 decoder & Baseline profile to MVC & JavaScript & 5 & S \\
  Java virtual machine & class-file versions 45 to 69 & C & 6 & S \\
  Kubernetes control plane & 1.25 to 1.33 & C & 9 & S \\
  Lua interpreter & Lua 5.1 to 5.5 & Java & 5 & S \\
  Matrix homeserver & Client-Server API v1.1 to v1.15 & C & 8 & S \\
  OpenType engine & TrueType and CFF to variable, color, and math fonts & JavaScript & 6 & S \\
  PDF library & PDF 1.3 to PDF 2.0, PDF/A, and PDF/UA & C & 8 & S \\
  PHP interpreter & PHP 5.6 to 8.4 & C & 11 & S \\
  PostgreSQL server & PostgreSQL 9.6 to 18 & C & 10 & S \\
  Protocol Buffers and gRPC & proto2 to editions 2024 & C & 8 & S \\
  Redis server & Redis 2.8 to 8.0 & Java & 11 & S \\
  Ruby interpreter & CRuby 1.8.7 to 3.4 & C & 6 & S \\
  SMB server & SMB1 to SMB 3.1.1 & JavaScript & 6 & S \\
\end{longtable}
}

% ===========================================================================
\section{Additional Details}
\label{app:details}

\paragraph{Agent harnesses.}
\Cref{tab:harness} lists the harness and reasoning effort for each model. Every
harness is used as provided by Harbor, without custom tools or prompts, and the
agent receives each step's \texttt{instruction.md} as its task. The
instructions, environments, and verifiers of all steps are included in the
released benchmark.

\begin{table}[H]
  \centering
  \small
  \begin{tabular}{@{}l l l@{}}
    \toprule
    Model & Harness & Reasoning effort \\
    \midrule
    Claude Opus 5.5     & Claude Code    & max \\
    Claude Fable 5.1    & Claude Code    & max \\
    Gemini 4 Argon      & mini-swe-agent & high \\
    GPT-6.1 Sol         & Codex CLI      & max \\
    GPT-6 Astra         & Codex CLI      & max \\
    DeepSeek V4.1 Flash & Claude Code    & max \\
    Qwen 3.8 Max        & Claude Code    & max \\
    MiMo 2.6 Pro        & Claude Code    & max \\
    Grok 4.7            & Grok Build     & xhigh \\
    \bottomrule
  \end{tabular}
  \caption{Agent harness and reasoning effort for each model.}
  \label{tab:harness}
\end{table}

\paragraph{Final reports.}
The analysis of final reports in \Cref{sec:result-analysis} uses a stratified
random sample of 2,157 steps: for each model, 140 to 143 steps that failed and 91
to 100 that passed. For each sampled step, Claude Opus 5.5 reads the agent's last messages,
without being told the model or the verifier's outcome, and assigns one of four
labels with a fixed rubric: the agent reports the step as complete and names no
shortcoming; it names a specific failing check, missing feature, or other
unresolved problem; it cites the time budget as a reason for stopping; or it
gives no final report. Shares are weighted by the number of passed and failed
steps of each model, and their 95\% bootstrap intervals span at most about 3
points on either side for the pooled shares. Labeling 209 of the steps a second
time, independently (218 label pairs), gives the same label in 94.5\% of pairs
(Cohen's $\kappa = 0.90$).

\paragraph{Reuse of upstream kernels.}
Each optimization task starts from a public library whose optimized kernels
were removed, so a model could in principle pass by recalling the upstream code
instead of writing its own. We check this on simdutf, whose verifier records
how much of the agent's code is upstream's: of the non-trivial lines that the
agent adds (excluding blank lines, braces, comments, and preprocessor
directives), it computes the fraction that appear verbatim anywhere in the
upstream source tree. Across the 252 simdutf steps in our runs, the median
fraction is 9\% and the maximum is 21\%. This is an upper bound, since ordinary
C++ lines also match upstream's by chance; counting only matches in the
corresponding upstream file, the median is 0.2\% and the maximum 5.7\%. A
verbatim-line measure does not detect copied code that has been reformatted or
renamed, so this result does not rule out memorization, but it indicates that
agents write most of their kernels themselves.

% ===========================================================================
\section{Raw Optimization Speedups}
\label{app:speedups}

The optimization reward in \Cref{eq:opt-score} is binary, so the step rewards
do not show how much faster an agent's code actually becomes. This appendix
reports the underlying raw speedups. For each optimization step, we take the
geometric-mean speedup $s$ recorded by the verifier; each per-function speedup
is floored at $1\times$ before averaging, so every correct record satisfies
$s \ge 1$. A step record is counted as correct if its output passes the
correctness gate. Quantiles use linear interpolation, and the geometric mean
is computed over correct records.

\paragraph{Overall.}
The 18 optimization tasks yield 4,068 step records across all models, of which
3,837 (94.3\%) pass the correctness gate. Over these correct records, the raw
speedup has a 25th percentile of $2.08\times$, a median of $3.71\times$, a 75th
percentile of $7.21\times$, a 90th percentile of $13.3\times$, a geometric mean
of $4.11\times$, and a maximum of $215\times$; 98.2\% of the correct records
exceed $1\times$. Correctness is therefore rarely the obstacle to full credit:
88.2\% of the optimization steps that fail to receive full credit produce
correct output but do not exceed the threshold $s_{\mathrm{ref}}$.

\paragraph{By model.}
\Cref{tab:speedup-models} reports the raw speedups of each model. The two
Claude models achieve the highest median and geometric-mean speedups, which is
consistent with their having the highest solve rates in the optimization
family (\Cref{tab:results}).

\begin{table}[H]
  \centering
  \small
  \begin{tabular}{@{}l r r r r r r@{}}
    \toprule
    Model & Correct & P50 & P75 & P90 & Geomean & Max \\
    \midrule
    Claude Opus 5.5     & 98.7 & 4.7 & 9.3 & 18.2 & 5.1 & 215 \\
    Claude Fable 5.1    & 99.6 & 4.7 & 10.2 & 16.0 & 5.5 & 208 \\
    Gemini 4 Argon      & 92.5 & 3.5 & 7.9 & 14.9 & 3.7 & 205 \\
    GPT-6.1 Sol         & 93.1 & 3.4 & 6.0 & 12.6 & 3.7 & 95.6 \\
    GPT-6 Astra         & 91.4 & 3.6 & 7.0 & 12.2 & 4.1 & 55.5 \\
    DeepSeek V4.1 Flash & 97.6 & 3.8 & 7.4 & 12.2 & 4.4 & 123 \\
    Qwen 3.8 Max        & 92.5 & 3.5 & 6.9 & 11.9 & 3.9 & 92.5 \\
    MiMo 2.6 Pro        & 97.1 & 3.0 & 6.1 & 10.3 & 3.5 & 42.9 \\
    Grok 4.7            & 86.5 & 2.9 & 5.8 & 11.7 & 3.4 & 82.8 \\
    \bottomrule
  \end{tabular}
  \caption{Raw optimization speedups by model, pooled over the 18 optimization
    tasks. Correct is the fraction of step records that pass the correctness
    gate (\%); the speedup statistics ($\times$) are computed over correct
    records.}
  \label{tab:speedup-models}
\end{table}

\paragraph{By task.}
\Cref{tab:speedups} reports, for each optimization task, the full-credit
threshold $s_{\mathrm{ref}}$ of every step and the distribution of raw speedups pooled
over all models and steps. The thresholds are the speedups of the reference
solutions as measured by the verifiers. Thresholds vary widely across
steps, from $1.2\times$ for the fourth LAMMPS step
to $37.6\times$ for the LDPC-decoding step of srsRAN, which reflects how much
faster than the portable code the reference solution is for each kernel family.

\begin{table}[H]
  \centering
  \footnotesize
  \setlength{\tabcolsep}{4pt}
  \begin{tabular}{@{}>{\raggedright\arraybackslash}p{3.5cm} p{4.9cm} r r r r r@{}}
  \toprule
  Task & $s_{\mathrm{ref}}$ per step ($\times$) & Correct & P50 & P90 & Geomean & Max \\
  \midrule
  Apache Arrow & 2.8, 2.3, 1.8, 1.7, 1.2 & 96 & 2.8 & 8.5 & 3.1 & 10.7 \\
  AviSynth+ & 2.8, 8.6, 2.8, 8.5, 2.9, 4.1, 7.0 & 96 & 4.4 & 9.5 & 4.2 & 16.5 \\
  BitMagic & 1.9, 7.8, 3.2, 3.1, 5.0, 1.6 & 99 & 3.7 & 11.7 & 4.1 & 21.1 \\
  BoringSSL & 24.7, 11.3, 5.6, 1.9, 2.3 & 89 & 5.8 & 25.6 & 5.1 & 51.3 \\
  FFmpeg libswscale & 9.0, 11.3, 4.9, 7.1, 6.9, 3.4 & 82 & 7.2 & 15.4 & 5.8 & 27.4 \\
  ISA-L crypto & 9.8, 3.7, 3.0, 6.0, 10.2, 5.0, 3.0 & 96 & 5.0 & 11.5 & 4.8 & 15.6 \\
  LAMMPS & 2.0, 3.2, 2.3, 1.2, 2.7 & 90 & 2.7 & 5.5 & 2.7 & 9.7 \\
  mldsa-native (ML-DSA) & 7.1, 4.1, 16.8, 15.7, 1.9, 5.1 & 99 & 5.0 & 19.2 & 5.8 & 24.3 \\
  mlkem-native (ML-KEM) & 1.5, 6.2, 11.4, 5.7, 8.9, 2.4, 10.9 & 96 & 4.5 & 12.1 & 4.2 & 18.5 \\
  OpenJPH (HTJ2K) & 1.7, 2.4, 1.8, 2.1, 4.7, 1.5 & 94 & 1.7 & 4.5 & 1.8 & 4.9 \\
  parasail & 4.0, 2.1, 2.7, 3.4, 1.8, 2.2, 2.5 & 95 & 3.0 & 5.3 & 2.7 & 11.0 \\
  Simd Library & 3.2, 18.8, 3.0, 3.4, 6.5, 2.9, 8.1, 7.6 & 90 & 4.0 & 11.0 & 4.2 & 21.7 \\
  simdutf & 1.9, 2.6, 1.8, 2.2, 2.5, 2.6, 2.3 & 97 & 2.3 & 3.0 & 2.2 & 3.8 \\
  srsRAN 5G PHY & 3.9, 4.5, 37.6, 10.1, 7.7, 11.4 & 99.5 & 8.1 & 40.3 & 9.4 & 215 \\
  StringZilla & 9.9, 8.9, 4.6, 2.6, 21.5, 24.0 & 95 & 12.5 & 40.7 & 11.1 & 81.7 \\
  SVT-JPEG-XS & 4.2, 2.0, 16.5, 1.9, 4.2, 2.2, 2.4, 6.9 & 95 & 4.9 & 18.8 & 5.6 & 60.2 \\
  uvg266 (VVC) & 1.7, 2.8, 1.9, 3.7, 2.3, 6.4 & 91 & 2.1 & 6.2 & 2.4 & 15.7 \\
  Vortex & 1.9, 2.1, 2.0, 4.3, 3.2 & 96 & 2.2 & 7.6 & 3.1 & 13.8 \\
  \bottomrule
  \end{tabular}
  \caption{Raw optimization speedups by task. $s_{\mathrm{ref}}$, the speedup
    of the reference solution, is listed per step, in step order. Correct is the fraction of step records that pass the
    correctness gate (\%); P50, P90, geometric mean, and maximum speedups
    ($\times$) are computed over correct records.}
  \label{tab:speedups}
\end{table}

\end{document}